\documentclass[11pt,twoside]{article}
\usepackage[T1]{fontenc}
\usepackage{graphicx}
\usepackage{color}
\usepackage{amsmath}
\usepackage[mathscr]{eucal}
\usepackage{amssymb}
\usepackage{graphicx}
\usepackage{graphicx,type1cm,eso-pic,color}
\usepackage{amssymb}
\usepackage{amssymb,amsmath}
\usepackage{graphicx,epsfig}
\usepackage{enumerate}
\usepackage{color}
\usepackage {multicol}

\numberwithin{equation}{section}
\newtheorem{theorem}{Theorem}
\newtheorem{lemma}{Lemma}

\newtheorem{remark}{Remark}

\newtheorem{example}{Example}

\makeatletter
\AddToShipoutPicture{
	\setlength{\@tempdimb}{.5\paperwidth}
	\setlength{\@tempdimc}{.5\paperheight}
	\setlength{\unitlength}{1pt}
	\put(\strip@pt\@tempdimb,\strip@pt\@tempdimc)
}
\makeatother

\begin{document}
	\setcounter{page}{1}

	\thispagestyle{empty}
	\markboth{}{}

	\pagestyle{myheadings}
	\markboth {On partial monotonicity of some extropy measures}{N. Gupta et al.}
	
	\date{}
	
	
	\noindent  
	
	\vspace{.1in}
	
	{\baselineskip 20truept
		
		\begin{center}
			{\Large {\bf On partial monotonicity of some extropy measures}} \footnote{\noindent	{\bf ** }  Corresponding author E-mail: skchaudhary1994@kgpian.iitkgp.ac.in}\\
			
			
	\end{center}}
	
	\vspace{.1in}
	
	\begin{center}
		{\large {\bf Nitin Gupta* and  Santosh Kumar Chaudhary**}}\\
		\vspace{.1in}
		{\large {{\bf *} Department of Mathematics, Indian Institute of Technology Kharagpur, Kharagpur 721302, West Bengal, India \\
				\bf nitin.gupta@maths.iitkgp.ac.in }} \\
		\vspace{0.2cm}
		{\large {{\bf **} Department of Mathematics, Indian Institute of Technology Kharagpur, Kharagpur 721302, West Bengal, India \\
				\bf skchaudhary1994@kgpian.iitkgp.ac.in}}\\
	\end{center}
	
	\vspace{.1in}
	\baselineskip 12truept

	\begin{abstract}
		\noindent Gupta and Chaudhary \cite{ngsk22a} introduced general weighted extropy and studied related properties. In this paper, we study conditional extropy and define the monotonic behaviour of conditional extropy. Also, we obtain results on the convolution of general weighted extropy.\\

		
	\end{abstract}
	
	\section{Introduction}
	In the technological age we live in, technology is a part of almost everything. In the field of computer science, the most well-known technology for allowing a computer to automatically learn from the past is called machine learning. Entropy and extropy in machine learning are two of the many techniques and concepts that are being used to solve complex problems easily. Further, entropy and extropy are also useful in the fields of information theory, physics, probability and statistics, computer science, economics, communication theory etc (see Balakrishnan et al. \cite{bala22},  Becerra et al. \cite{becerra2018},  Kazemi et al. \cite{kazemi21fde}, Sati and Gupta \cite{satigupta2015}, Tahmasebi and Toomaj \cite{tahmasebi22toomaj}, Tuli \cite{tuli2010}).
	
	Shannon \cite{shannon1948} introduced the notion of information entropy which measures the average amount of uncertainty about an occurrence associated with a certain probability distribution. Let $Y$ be a discrete random variable having probability mass function $p_i,\ i=1,2\ldots, N$. The discrete version of Shannon entropy  is defined, as	
	\[H_N(Y) =-{\sum\limits_{i = 1}^{N} {p_{i} \log\left({p_{i}}\right)}}.\]
	Here and throughout this paper, $\log$ denotes logarithm to base $e$. Let $Y$ be an absolutely continuous random variable having probability density function $g_{Y} (y)$ and cumulative distribution function $G_{Y}(y)$. This notation of probability density function and cumulative distribution function will be carried out throughout the paper. The differential form of Shannon entropy  is defined, as
	\begin{equation*}\label{1eq2}
		H(Y)=-{\int_{-\infty}^{\infty}{g_Y(y) \log\left({g_Y(y)}\right)dy}}.
	\end{equation*} 
	Various researchers have suggested different generalisations of entropy to measure uncertainty (see Gupta and Chaudhary \cite{ngsk22a}, Hooda \cite{hooda2001}, Jose and Sathar \cite{josesathar19}, Kayal and Vellaisamy \cite{kayalvelai2011gep}, Kayal \cite{kayal21failureex}, Qiu \cite{qiujia18}, Sathar and Nair \cite{satharnair19}). 
	
	Cumulative past entropy was proposed and studied by Di Crescenzo and Longobardi \cite{dicresmlong2009} as
	\begin{equation*}\label{CPE}
		\xi (Y)=-{\int_{-\infty}^{\infty}{G_Y(y) \log \left({G_Y(y)}\right)dy}}.
	\end{equation*}
	Conditional cumulative past entropy of $Y$ given $S=(c,d)$ is given as 
	\begin{equation*}\label{CPE}
		\xi (Y|S)=-{\int_{-\infty}^{\infty}{G_{Y|S}(y) \log \left({G_{Y|S}(y)}\right)dy}}.
	\end{equation*}
	
	R\`enyi \cite{renyi1961} proposed the generalized entropy of order $\theta$ for $\theta > 0, \,\theta \ne 1,$ which is given by
	\begin{equation*}
		\label{renyi}
		H_\theta(Y)=\frac{1}{1-\theta}\log\left(\int_{-\infty}^{\infty}(g_Y(y))^\theta dy\right).
	\end{equation*}
	Tsallis \cite{tsallis1988}  defined the generalized entropy for $\theta > 0, \,\theta \ne 1,$ which is given by 
	\begin{equation*}
		\label{tsallis}
		S_\theta(Y)=\frac{1}{\theta-1}\left(1-\int_{-\infty}^{\infty}(g_Y(y))^\theta dy\right).
	\end{equation*}
	Kapur \cite{kapur1967} gives Kapur entropy of order $\theta$ and type $ \lambda $ for $\theta\neq \lambda ,\,\,\theta >0,\,\lambda >0,$  which is given by
	\begin{equation*}
		H_{\theta,\lambda} (Y)=\frac{1}{\lambda-\theta}\left[ \log\left(\int_{-\infty}^{\infty}(g_Y(y))^\theta dy\right)-\log\left({\int_{-\infty}^{\infty}(g_Y(y))^\lambda dy}\right) \right].
	\end{equation*}
	Varma \cite{varma1966}  generalized entropy of order $\theta$ and type $\lambda$ for $\lambda - 1 < \theta < \lambda ,\,\,\,\lambda \ge 1,$ which is given by
	\begin{align*}
		\label{verma}
		H_{\theta} ^{\lambda}  (Y) = {\frac{{1}}{{\lambda - \theta}} }\log \left({{\int_{-\infty}^{\infty}  {\left( {g_{Y} (y)} \right)^{\theta + \lambda -1}}} dy} \right).
	\end{align*}

	The conditional Shannon entropy of $Y$ given $S$, where $S=\{c<Y<d\}$, is given by
	\[
	H\left( Y|S \right) = -\int_c^d g_{Y|S}(y)\log \left(g_{Y|S}(y)\right) dy,
	\]
	where \[g_{Y|S}(y) = {\frac{{g_{Y} (y)}}{{G_{Y}(d)-G_{Y}(c)}}}\, ,\,\ c<y<d .\]
	
	One may refer to Sunoj et al. \cite{sunojetal2009}  for a review of conditional Shannon entropy ($H\left( Y|S \right)$). Convolution and monotonic behaviour of the conditional Shannon entropy, Renyi entropy, Tsallis entropy, Kapur's and Verma's entropies have been studied in the literature (see Chen et al. \cite{chenetal2010}, Gupta and Bajaj \cite{guptabajaj2013}, Sati and Gupta \cite{satigupta2015} and Shangari and Chen \cite{shangri2012}). Bansal and Gupta \cite{bansalgupta2020conf} studied the monotonicity properties of conditional cumulative past entropy $\xi (Y|S)$ and convolution results for conditional extropy $J(Y|S)$. In this paper, we study the monotonicity of conditional extropy and convolution results for general weighted extropy.
	
	As described in Chen et al. \cite{chenetal2010} and Shangari and Chen \cite{shangri2012}, $H\left( Y|S \right)$ may serve as an indicator of uncertainty for an interval $S$. The measure of uncertainty shrinks/expands as the interval providing the information about the outcome shrinks/expands.  For intervals $S_1$ and $S_2$ such that  $S_2 \subseteq S_1$, then entropy $H$ is partially increasing (decreasing) if $H(Y|Y\in S_2)\le (\ge) H(Y|Y\in S_1)$. Under the condition that $G_Y\left( {y} \right)$ is a log-concave function (for more on log-concave probability and its application, see Bagnoli and  Bergstrom \cite{bagnoli2005berg}). Shangari and Chen \cite{shangri2012} proved that $H(Y|Y\in S)$ is a partially increasing function in the interval $S$. Under the same condition, they also proved that the conditional	Renyi entropy $H_\theta (Y|S)$ of $Y$ given $S =(c,d)$ is a partially increasing function in the interval $S$ for $\theta \ge	0\, \,\theta \ne 1$. Under the condition that  $G_Y\left( {y} \right)$ is concave, Gupta and Bajaj \cite{guptabajaj2013} proved that conditional Kapur entropy $H_{\theta, \lambda}(Y|S)$ of $Y$ given $S = (c,d)$ is a partially increasing function in the interval $S$. They also show that if $G_Y\left( {y} \right)$ is a log-concave function then the conditional Tsallis entropy $S_\theta (Y|S)$ of	$Y$ given $S$ is a partially increasing function in the interval $S$ where $S =(c,d)$. Sati and Gupta \cite{satigupta2015} studied the monotonic behaviour of conditional Varma entropy $H_{\theta} ^{\lambda}(Y|S)$. Under the condition $G_{Y} (y)$ is log-concave function and $\theta +\lambda > ( < )2$ they showed that the $H_{\theta} ^{\lambda}(Y|S)$ is partially decreasing (increasing) in $S=(c,d)$.

	Ash \cite{ash1990}, Cover and Thomas \cite{coverthomas2006} and Yeung \cite{yeung2002} provide an excellent review of detailed properties which play an important role in information theory. Bansal and Gupta \cite{bansalgupta2020conf} proposed a new conditional entropy which is based on cumulative past entropy and defined the monotonic behaviour of $\xi (Y|S)$. They proved that if $\int_{c}^{y}G_Y(u)du$ is a log-concave function then the conditional	cumulative past entropy $Y$ given $S$, i,e. $\xi (Y|S)$ is increasing in $d$ where $S=(c,d).$
	\newline \indent Entropies are significant in the study of likelihood bases, inference principles, and large deviation theory because of its relevance in these fields. Shannon, Renyi, Tsallis and  Varma entropies have operational meaning in terms of data compression. They also find a role as a measure of complexity and uncertainty in different areas such as coding theory, computer science, electronics and physics. For more details, one may refer to Cover and Thomas \cite{coverthomas2006}.
	\newline \indent Extropy, an alternative measure of uncertainty, was defined by Lad et al. \cite{ladf15ext} and proved to be the complement dual of the Shannon entropy. Entropy and extropy measures may be thought of as the positive and negative images of a photographic film related to each other. Extropy and its generalisations have numerous applications in the literature, including information theory, economics, communication theory, computer science, and physics see Balakrishnan et al. \cite{bala22}, Becerra et al. \cite{becerra2018}, Kazemi et al. \cite{kazemi21fde}, Tahmasebi and Toomaj \cite{tahmasebi22kazemi}).  Becerra et al. \cite{becerra2018} used extropy in speech recognition that can also be used to score the forecasting distribution. Balakrishnan et al. \cite{bala22}  used Tsallis extropy in pattern recognition. Based on a generalisation of extropy known as negative cumulative extropy, Tahmasebi and Toomaj \cite{tahmasebi22toomaj} investigated the stock market in OECD nations. The fractional Deng extropy, a generalisation of extropy, was investigated by Kazemi et al. \cite{kazemi21fde}  in relation to a classification problem. To solve the compressive sensing problem, Tahmasebi et al. \cite{tahmasebi22kazemi} applied certain extropy measures. Extropy provides some conceptual advantages over entropy in particular circumstances, despite the mathematical similarities between entropy and extropy. Extropy of  discrete random variable $Y$ with probability mass function $p_i$ for $i=1,2,\dots N $ is defined as	\[J_N(Y)=-\sum_{i=1}^{N}(1-p_i)\log(1-p_i).\] 
	The entropy and extropy are identical for $N=2$, that is, $J_2(Y)=H_2(Y)$. The extropy also called the differential extropy of continuous distribution with probability density function $g_Y(y)$ is 
	\[J(Y)=-\frac{1}{2}\int_{-\infty}^{\infty}g_Y^2(y)dy.\]
	The conditional extropy of $Y$ given $S=(c,d)$ is given as
	\begin{align*}
		J (Y|S)&=-\frac{1}{2} {\int_{-\infty}^{\infty}{g^2_{Y|S}(y)dy}}.
	\end{align*}

	Qiu \cite{qiu2017exorderstats} studied the extropy for order statistics and record values, including characterisation results, lower bounds, monotone properties, and statistical applications. Balakrishnan et al. \cite{balabuono2020} and Bansal and Gupta \cite{bansalgupta2021cis} independently introduced the weighted extropy as
	\[J^y(Y)=-\frac{1}{2}\int_{-\infty}^{\infty}yg_Y^2(y)dy.\]
	The conditional weighted extropy of $Y$ given $S=(c,d)$ is given as
	\begin{align*}
		J^y (Y|S)&=-\frac{1}{2} {\int_{-\infty}^{\infty}{yg^2_{Y|S}(y)dy}}.
	\end{align*}	
	The general weighted extropy of $Y$ with weight $w(y)\geq0$ is given as
	\begin{align*}
		J^w (Y)&=-\frac{1}{2} {\int_{-\infty}^{\infty}{w(y)g^2_{Y}(y) dy}}.
	\end{align*}
	The conditional general weighted extropy of $Y$ with weight $w(y)\geq0$ given $S=(c,d)$ is given as
	\begin{align*}
		J^w (Y|S)&=-\frac{1}{2} {\int_{-\infty}^{\infty}{w(y)g^2_{Y|S}(y)dy}}.
	\end{align*}	
	The $J^w (Y)$ is the generalization of weighted $J(Y).$	Balakrishnan et al. \cite{balabuono2020} studied characterization results and bounds for weighted versions of extropy, residual extropy, past extropy, bivariate extropy	and bivariate weighted extropy whereas Bansal and Gupta \cite{bansalgupta2021cis} discussed the results for weighted extropy and weighted residual extropy of $Y$. Gupta and Chaudhary \cite{ngsk22a} defined $J^w (Y)$ and provided some results related to ranked set sampling. Bansal and Gupta \cite{bansalgupta2020conf} studied $\xi (Y|S)$ and its partial monotonicity. This motivated us to find partial monotonicity of conditional extropy.  Further, convolution of extropy has also been studied by Bansal and Gupta \cite{bansalgupta2020conf}. That motivated us to derive results on convolution of general weighted extropy. This paper is arranged as follows. Section \ref{s1} provides the result on the monotonicity of conditional extropy. In Section \ref{s2}, we studied convolution of general weighted extropy. Section \ref{s3} concludes this paper.

	\section{ Monotonicity of conditional extropy}\label{s1}
	The conditional extropy of $Y$ given $S=(c,d)$ is given as
	\begin{align*}
		J (Y|S)&=-\frac{1}{2} {\int_{-\infty}^{\infty}{g^2_{Y|S}(y) dy}} \\
		&=-\frac{1}{2}{\int_{c}^{d}{ \left(\frac{g_Y(y)}{G_Y(d)-G_Y(c)} \right)^2dy}}.
	\end{align*}
	
	The measure of uncertainty shrinks/expands as the interval providing the information about the outcome shrinks/expands. Here we study the conditions under which the $J(Y|S)$ is increasing in $d$, where $S=(c,d)$ and that is given by the following theorem.
	\begin{theorem}\label{thm1}
		Let $S=\{c<Y<d\}$. If $G_Y(y)$ is log-concave (log-convex), then 
		$J (Y|S)$ is increasing (decreasing) in $d$, for fixed $c$.
	\end{theorem}
	{\bf Proof :} From definition,
	\begin{align*}
		J(Y|S)&=\frac{-1}{2} {\int_{c}^{d}{\left(\frac{g_Y(y)}{G_Y(d)-G_Y(c)} \right)^2dy}}.
	\end{align*}
	Now for fixed $c$, differentiating $J(Y|S)$ with respect to $d$, we get
	\begin{align}\label{thm1eq10}
		\frac{\text{d}(J(Y|S))}{\text{d} d}&=\frac{-1}{2\left(G_Y(d)-G_Y(c)\right)^4}\left(g_Y^2(d) \left( G_Y(d)-G_Y(c)\right)^2-2g_Y(d)\right.\nonumber\\
		&\ \ \ \ \  \ \ \ \ \ \ \  \ \ \ \ \ \ \ \ \ \  \ \ \ \ \ \  \ \ \ \ \  \ \ \ \ \ \  \left.\left(G_Y(d)-G_Y(c)\right)\int_{c}^{b}g_Y^2(y)dy\right)\nonumber\\
		=&\frac{g_Y(d)\psi_1(d)}{2\left(G_Y(d)-G_Y(c)\right)^3},
	\end{align}
	where
	\[\psi_1(y)=2 \int_{c}^{y}g_Y^2(u)du-g_Y(y)\left(G_Y(y)-G_Y(c)\right),\ \ \psi_1(c)=0,\]
	and
	\begin{align}\label{thm1eq11}
		\psi_1^\prime(y)&=g_Y^2(y)-g_Y^\prime(y)\left(G_Y(y)-G_Y(c)\right)\nonumber .
	\end{align}
	Note that $G_Y(y)$ is log-concave function, implies that $\frac{G_Y(y)-G_Y(c)}{G_Y(d)-G_Y(c)}$ is log-concave function. Hence we have
	\[g_Y^2(y)-g_Y^\prime(y)\left(G_Y(y)-G_Y(c)\right)\geq 0,\  \text{for all } d. \]
	Since $\psi_1^\prime(y)\geq 0$, that is, $\psi_1(y)$ is increasing in $y$. Now for $d>c$ we have $\psi_1(d)\geq  \psi_1(c)$, that is, $\psi_1(d)\geq  0$. Hence from (\ref{thm1eq10}) we have, $\frac{\text{d} J(Y|S)}{\text{d} d}\geq 0$. Therefore $J(Y|S)$ is increasing in $d$, for fixed $c$.
	\newline \indent In the next section, we provide a result on convolution of $J^w (Y)$. We will prove that the conditional general weighted extropy of $V = |Y_{1} - Y_{2}|$ given $S = \{c \le Y_{1}, \ Y_{2} \le d\}$, i,e. $J^w(V|S)$ is partially increasing in $S$.
	
	\section{ Convolution of General weighted extropy}\label{s2}
	Let $Y$ be a random experiment and it is repeated to measure its reproducibility or precision or both.  Then measure of uncertainty of the experiment is the function $V = |Y_{1} - Y_{2}| $; where $Y_{1}$ and  $Y_{2} $ are independent and identically distributed random variable from an experiment $Y$ with probability density function $g_Y(y)$. The difference $V=|Y_1-Y_2|$ is the measure of the uncertainty between two outcomes. Uncertainty should reduce if further information of the form  $S=\{c<Y_1, Y_2<d\}$ is provided.

	The marginal probability density function of $V=|Y_1-Y_2|$ given $S=\{c<Y_1, Y_2<d\}$  is	\[h(v;c,d) =  \int\limits_{c+v}^{d} {\frac{g_Y(y-v)g_Y(y) dy}{(G_Y(d)-G_Y(c))^2}}, \ \text{for all} \ v\in[0,d-c].\]

	Chen et al. \cite{chen2013} proved that the $H(V|Y\in S)$ is partially monotonic in $S$ provided the random variables $Y_1$ and $Y_2$ have log-concave probability density functions that take value in $S$. Shangari and Chen \cite{shangri2012} claimed and Gupta and Bajaj \cite{guptabajaj2013} proved that if $Y_1$ and $Y_2$ have log-concave probability density function which takes value in $S$, then the conditional Tasalli and Renyi entropy of $V$ given $S$ is partially increasing function in $S$ if $\theta>0$, $\theta\ne 1$. Sati and Gupta \cite{satigupta2015} study the partial monotonicity of the $H_{\theta} ^{\lambda}(Y|S)$.  Bansal and Gupta \cite{bansalgupta2020conf} studied the convolution results for conditional extropy.

	The proof of the next result of this section will be using the following lemma from Chen et al. \cite{chenetal2010} (also see Sati and Gupta \cite{satigupta2015}).
	\begin{lemma}\label{Chan}
		\begin{enumerate}[(a)]
			
			\item \label{Chen (2010)}
			Let the probability density functions of random variables $Y_1$ and $Y_2$ be log-concave functions. If the function $\phi(v)$ is increasing in $v$, then $E(\phi(V)|S)$ is increasing in $d$ for any $c$, and decreasing in $c$ for any $d$; where $V=|Y_1-Y_2|$ where $S=\{c<Y_1,Y_2<d\}.$
			\item \label{Chen (2010)a}
			If $g_Y(y)$ is log-concave function, then $h(v;c,d)$ is decreasing function of $v$ on $v\in [0,d-c]$.
		\end{enumerate}
	\end{lemma}

	Now, we will prove the following theorem which provides the conditions for $J^w (V|S)$ to be a partially increasing/decreasing in $S$.
	
	\begin{theorem}\label{sec3thm1} Let the probability density functions of random variables $Y_1$ and $Y_2$ be log-concave function. Let weight be $w(y)\geq 0$, $w(y)$ is decreasing in $y$, and $S=\{c<Y_1, Y_2<d\}$, then the $J^w (Y|S)$ is a partially increasing in $S$.
	\end{theorem}
	\noindent {\bf Proof :} The conditional general weighted extropy of  $V$ given $S$ is 
	\[J^w(V|S)=\frac{-1}{2} \int_{c}^{d}w(v)\left( h(v;c,d)\right)^{2}dv. \]
	For fixed $c$, if we choose for any $d_1\leq d_2$, \[\psi_1(v)=\left(w(v)\right)^{1/2}h(v;c,d_1)\ \ \text{and}\ \ \psi_2(v)=\left(w(v)\right)^{1/2}h(v;c,d_2)\] clearly here $\psi_1(v)$ and $\psi_2(v)$ are non-negative functions.
	Also, let $p=2$, $q=2$, then $p>0,\ q>0$ and $\frac{1}{p}+\frac{1}{q}=1$. With the help of H$\ddot{o}$lder's inequality, we now obtain
	\begin{align}\label{sec3eq1}
		&\int \psi_1(v)\psi_2(v)dv\leq \left(\int(\psi_1(v))^p dv \right)^{1/p}\left(\int (\psi_2(v))^q dv\right)^{1/q},\nonumber\\
		& \text{that is,}\ \ \int w(v)h(v;c,d_1)h(v;c,d_2)dv\leq \left(\int w(v) h^2(v;c,d_1) dv \right)^{1/2}\nonumber\\
		& \ \ \ \ \ \  \ \ \ \ \ \ \ \ \ \ \ \ \  \ \ \ \ \ \ \  \ \ \ \ \ \  \ \ \ \ \ \ \  \ \ \ \ \ \  \ \ \ \ \ \ \  \ 
		\left(\int w(v) h^2(v;c,d_2) dv\right)^{1/2}.
	\end{align}
	For fixed $d>0$, let
	\[\phi_1 (v)=-w(v)h(v;c,d) ;\]
	then,
	\[\phi_1^\prime(v)=-w^{\prime}(v)h(v;c,d)-w(v)h^{\prime}(v;c,d)\geq 0\]
	as the probability density function $h(v;c,d)$ is decreasing function in $v$ for $0\leq v\leq d-c$ (Using Lemma \ref{Chan} (\ref{Chen (2010)a})), $w(v)\geq 0$, and $w(v)$ is decreasing function in $v$ for $0\leq v\leq d-c.$ 
	Hence $\phi_1 (v)$ increases in $v$. Therefore, by lemma \ref{Chan} (\ref{Chen (2010)}) for any $c<d_1<d_2$, we have
	\begin{align}
		E(\phi_1 (V)|c\leq Y_1,Y_2 \leq d_1)&\leq E(\phi_1 (V)|c\leq Y_1,Y_2 \leq d_2),\nonumber\\
		\text{that is,}\ \ \int w(v) h^2(v;c,d_2)dv&\leq \int w(v)h(v;c,d_1) h(v;c,d_2)dv.  \label{sec3eq2}
	\end{align}
	From (\ref{sec3eq1}) and (\ref{sec3eq2}), we have
	\begin{align}\label{sec3eq3}
		\int w(v) h^2(v;c,d_2)dv &\leq \int w(v) g^2(v;c,d_1)dv.
	\end{align}
	Therefore we have
	\begin{align*}
		-\frac{1}{2}\int w(v) h^2(v;c,d_1)dv &\leq  -\frac{1}{2} \int   w(v) h^2(v;c,d_2)dv,\\
		\text{that is,},\  J^w(V|c<Y_1,Y_2<d_1)&\leq J^w(V|c<Y_1,Y_2<d_2); \text{for}\ d_1\le d_2.
	\end{align*}
	Hence, for a fixed $c$, $J^w(V|S)$ is increasing in $d$.

	Now for fixed $d$, if we choose for any $c_1\leq c_2$,
	\[\psi_3(v)=w^2(v) h^2(v;c_1,d)\ h^2(v;c_2,d) \ \ \text{and} \ \  \psi_4(v)=\left(w(v)h^2(v;c_1,d)\right)^{-1}.\] Clearly $\psi_3(v)$ and $\psi_4(v)$ are non-negative.
	Also, let $p=\frac{1}{2}$, $q=-1$, then $p<1,\ q<0$ and $\frac{1}{p}+\frac{1}{q}=1$. Now H$\ddot{o}$lder's inequality provides
	\begin{align}\label{sec3eq3b}
		&\left(\int(\psi_3(v))^p dv \right)^{1/p}\left(\int (\psi_4(v))^q dv\right)^{1/q}\leq \int \psi_3(v)\psi_4(v)dv, \nonumber\\
		&\text{that is,}\  \left(\int w(v) h(v;c_1,d) \ h(v;c_2,d)dv\right)^{2}\left(\int w(v)(h(v;c_1,d))^{2}dv \right)^{-1}\nonumber\\ & \ \ \ \ \ \ \ \ \ \ \ \ \ \ \ \ \ \ \ \ \ \  \ \ \ \ \ \ \ \ \ \ \ \ \ \  \ \ \ \ \ \ \ \ \ \ \ \ \ \ \ \ \leq \int w(v)(h(v;c_2,d))^{2}dv, \nonumber\\
		&\text{that is,}\  \int w(v) h(v;c_1,d)\ h(v;c_2,d)dv \leq \left(\int w(v) (h(v;c_1,d))^{2}dv\right)^{\frac{1}{2}}\nonumber\\
		& \ \ \ \ \ \  \ \ \ \ \ \ \ \ \ \ \ \ \  \ \ \ \ \ \ \  \ \ \ \ \ \  \ \ \ \ \ \ \  \ \ \ \ \ \  \ \ \ \ \ \ \  \ \left(\int w(u) (h(v;c_2,d))^{2}dv\right)^{\frac{1}{2}}.
	\end{align}
	For fixed $c_2>0$, let
	\[\phi_2 (v)=-w(v)h(v;c_1,d) ;\]
	then,
	\[\phi_2^\prime(v)=-w^\prime (v)h(v;c_1,d)-w(v)h^\prime(v;c_1,d)\geq 0,\]
	as the probability density function $h(v;c,d)$ is decreasing function in $v$ for $0\leq v\leq d-c$ (Using Lemma \ref{Chan} (\ref{Chen (2010)a})), $w(v)\geq 0$, and $w(v)$ is decreasing function in $v$.
	Hence $\phi_2 (v)$ increases in $v$. By lemma \ref{Chan} (\ref{Chen (2010)}) for any $c_1<c_2<d$, we have
	\begin{align}
		E(\phi_2 (V)|c_2\leq Y_1,Y_2 \leq d)&\leq E(\phi_2 (V)|c_1\leq Y_1,Y_2 \leq d),\nonumber\\
		\text{that is,}\ \int w(v) \ (h(v;c_1,d))^{2}dv&\leq  \int w(v) \  h(v;c_1,d)h(v;c_2,d)dv.\label{sec3eq4}
	\end{align}
	Now, (\ref{sec3eq3}) and (\ref{sec3eq4}) implies
	\begin{align}\label{sec3eq5}
		\int w(v) \ (h(v;c_1,d))^{2}dv \le \int  w(v) \ (h(v;c_2,d))^{2}dv.
	\end{align}
	Therefore we have
	\begin{align*}
		-\frac{1}{2}\int w(v) \ (h(v;c_2,d))^{2}dv &\le -\frac{1}{2}\int w(v) \  (h(v;a_1,d))^{2}dv,\\
		\text{that is,}\ \ J^w(V|c_2<Y_1,Y_2<d)&\leq J^w(V|c_1<Y_1,Y_2<d);\ \ \text{for}\ c_1\le c_2.
	\end{align*}
	As a result, for fixed $d$, the $J^w(V|S)$ is decreasing in $c$. Therefore the $J^w(V|S)$ is partially increasing in $S$.\\

	\noindent \textbf{Remarks:}
	It is observable from the above theorem that, under specific circumstances, the $J(Y|S)$ is partially increasing in $S$, demonstrating its reasonability as a complement dual of the entropy measure.

	\begin{remark}
		In  Theorem \ref{sec3thm1} if we take $w(y)=1$, we get the result of Bansal and Gupta \cite{bansalgupta2020conf}.
	\end{remark}

	The following examples to Theorem \ref{sec3thm1} may be provided.
	\begin{example}
		\begin{enumerate}[(a)]
			\item Let $Y_1$ and $Y_2$ be two independent and identically distributed Weibull random variables with probability density function for $\theta\ge 1,\ \lambda \ge 0$,
			\[g_Y(y)=\theta \lambda^{\theta} y^{\theta-1}e^{-(\lambda y)^{\theta}},\ y\ge 0.\]
			Since $g_Y(y)$ is a log-concave function for $\theta \ge 1$ and  $w(y)=1/y$, then using Theorem \ref{sec3thm1}, $J^w (V|S)$ is a partially increasing in $S$.
			
			\item Let $Y_1$ and $Y_2$ be two independent and identically distributed gamma random variables with probability density function for $\theta \ge 1,\  \lambda \ge 0$,
			\[g_Y(y)=\frac{\lambda^\theta }{\Gamma (\theta)}y^{\theta-1}e^{-\lambda y},\ \ y\ge 0.\]
			Since the probability density function of the gamma distribution is a log-concave function for $\theta \ge 1$ and let $w(y)=1/y$,
			then using Theorem \ref{sec3thm1}, $J^w (V|S)$ is a partially increasing in $S$. .
		\end{enumerate}
	\end{example}
	
	\section{Conclusion} \label{s3}
	The extropy measure and its generalisations are now widely used in all scientific domains. General weighted extropy is a generalisation of extropy. We proposed conditional extropy and studied its partial monotonicity. We also obtained some results on convolution of general weighted extropy.\\
	
	\noindent \textbf{\Large Funding} \\
	\\
	Santosh Kumar Chaudhary is getting financial assistance for research from the Council of Scientific and Industrial Research (CSIR), Government of India (File Number 09/0081 (14002)/2022-EMR-I).\\
	\\
	\noindent \textbf{\Large Conflict of interest} \\
	\\
	No conflicts of interest are disclosed by the authors.\\
	
	\noindent \textbf{\Large Acknowledgement} \\
	\\
	The authors are thankful to the reviewers for their insightful comments, which significantly improved this manuscript.


\begin{thebibliography}{99}
		
		\bibitem{ash1990} Ash, R.B.: Information Theory. Dover Publications Inc., New York (1990)
		
		\bibitem{bagnoli2005berg} Bagnoli, M., Bergstrom, T.: Log-concave probability and its applications. Econ.	Theory 26(2), 445–469 (2005)
		
		\bibitem{bala22} Balakrishnan, N., Buono, F., Longobardi, M.: On Tsallis extropy with an application to pattern recognition. Statistics \& Probability Letters. 180, 109241 (2022)
		
		\bibitem{balabuono2020} Balakrishnan, N., Buono, F., Longobardi, M.: On weighted extropies. Communications in Statistics-Theory and Methods. 1-31 (2020) DOI: 10:1080=03610926:2020:1860222
		
		\bibitem{bansalgupta2021cis} Bansal, S.,  Gupta, N.:  Weighted extropies and past extropy of order statistics and k-record values. Communications in Statistics - Theory and Methods. 1-18 (2021)
		
		\bibitem{bansalgupta2020conf} Bansal, S., Gupta, N.: On Partial Monotonic Behaviour of Past Entropy and Convolution of Extropy. In: Castillo, O., Jana, D., Giri, D., Ahmed, A. (eds) Recent Advances in Intelligent Information Systems and Applied Mathematics. ICITAM 2019. Studies in Computational Intelligence. vol 863. Springer, Cham (2020)  https://doi.org/10.1007/978-3-030-34152-7\_16
		
		\bibitem{becerra2018} Becerra, A., de la Rosa, J.I., González, E., Escalante,  N. I.: Training deep neural networks with non-uniform frame-level cost function for automatic speech recognition. Multimed Tools Appl. 77, 27231–27267 (2018)
		
		\bibitem{chen2013} Chen, J.: A partial order on uncertainty and information. J. Theor. Probab. 26(2), 349–359 (2013)
		
		\bibitem{chenetal2010} Chen, J., van Eeden, C., Zidek, J.V.: Uncertainty and the conditional variance. Stat. Probab. Lett. 80,	1764–1770 (2010)
		
		\bibitem{coverthomas2006} Cover, T., Thomas, J.A.,: Elements of Information Theory. second ed. John Wiley \& Sons Inc., Hoboken, NJ (2006)
		
		
		\bibitem{dicresmlong2009} Di Crescenzo, A., Longobardi, M.: On cumulative entropies. J. Stat. Plann. Infer. 139(12), 4072–4087 (2009)
		
		\bibitem{ladf15ext} Lad, F., Sanfilippo, G., Agro, G.: Extropy: complementary dual of entropy. Stat. Sci. 30(1), 40–58 (2015)
		
		\bibitem{guptabajaj2013} Gupta, N., Bajaj, R.K.: On partial monotonic behaviour of some entropy measures. Stat. Probab. Lett. 83(5), 1330–1338 (2013)
		
		\bibitem{ngsk22a} Gupta, N., Chaudhary, S.K.: On general weighted extropy of ranked set sampling. arXiv preprint (2022)  doi: 10.48550/ARXIV.2207.02003 (Communicated to journal)
		
		\bibitem{hooda2001} Hooda, D.: A coding theorem on generalized r-norm entropy. Korean J. Comput. Appl. Math. 8(3), 657–664 (2001)
		
		\bibitem{josesathar19} Jose, J., Abdul Sathar, E.: Residual extropy of k-record values. Stat. Probab. Lett. 146, 1–6 (2019)
		
		\bibitem{kazemi21fde} Kazemi, M.R., Tahmasebi, S., Buono, F., Longobardi, M.: Fractional Deng Entropy and Extropy and Some Applications. Entropy. 23, 623 (2021)
		
		\bibitem{kapur1967} Kapur, J.N.: Generalized entropy of order $\alpha$ and type $\beta$. In: The Math. Seminar, vol. 4, pp. 78–82 (1967)
		
		\bibitem{kayalvelai2011gep} Kayal, S., Vellaisamy, P.:  Generalized entropy properties of records. Journal of Analysis. 19, 25-40 (2011)
		
		\bibitem{kayal21failureex} Kayal, S.:  Failure extropy, dynamic failure extropy and their weighted versions. Stochastics and Quality Control. 36(1), 59-71 (2021)
		
		\bibitem{qiu2017exorderstats} Qiu, G.: The extropy of order statistics and record values. Stat. Probab. Lett. 120, 52–60 (2017)
		
		\bibitem{qiujia18} Qiu, G., Jia, K.: The residual extropy of order statistics. Stat. Probab. Lett. 133, 15–22 (2018)
		
		\bibitem{renyi1961} R\`enyi, A.: On measures of entropy and information. Tech. rep, Hungarian Academy of Sciences Budapest Hungary (1961)
		
		\bibitem{satharnair19} Sathar, E. I. A., Nair, R.D.:  On dynamic survival extropy, Communications in Statistics - Theory and Methods. 50(6), 1295-1313 (2019) 
		
		\bibitem{satigupta2015} Sati, M. M., Gupta, N.: On partial monotonic behaviour of Varma entropy and its application in coding theory. Journal of the Indian Statistical Association. 53, 135-152 (2015)
		
		\bibitem{shangri2012} Shangri, D., Chen, J.: Partial monotonicity of entropy measures. Statistics and Probability Letters. 82(11), 1935-1940 (2012)
		
		\bibitem{shannon1948} Shannon, C.E.: A mathematical theory of communication. Bell System Technical Journal. 27, 379-423, 623-656 (1948)
		
		
		\bibitem{sunojetal2009} Sunoj, S., Sankaran, P., Maya, S.: Characterizations of life distributions using conditional expectations of doubly (interval) truncated random variables. Commun. Stat.-Theory Methods. 38(9), 1441–1452 (2009).
		
		\bibitem{tahmasebi22toomaj} Tahmasebi, S., Toomaj, A.: On negative cumulative extropy with applications. Commun. Stat. Theory Methods. 51, 5025–5047 (2022)
		
		\bibitem{tahmasebi22kazemi} Tahmasebi, S., Kazemi, M.R., Keshavarz, A., Jafari, A.A., Buono, F.: Compressive Sensing Using Extropy Measures of Ranked Set Sampling. Math. Slovaca, accepted for publication (2022) 
		
		\bibitem{tsallis1988} Tsallis, C.: Possible generalization of Boltzmann-Gibbs statistics. J. Stat. Phys. 52, 479–487 (1988)
		
		\bibitem{tuli2010} Tuli, R.: Mean codeword lengths and their correspondence with entropy measures. Int. J. Eng. Nat. Sci. 4, 175–180 (2010)
		
		\bibitem{varma1966} Varma, R.: Generalizations of Renyi’s entropy of order $\alpha$. J. Math. Sci. 1, 34–48 (1966)
		
		\bibitem{yeung2002} Yeung, R. W.: A First Course in Information Theory. Kluwer Academic/Plenum Publishers. New York (2002)
		
	\end{thebibliography}
\end{document}